\documentclass[cits]{PoS}

\pdfoutput=1

\usepackage{verbatim,mathbbol}

\def\tr{\mbox{tr }}

\title{How instantons could survive the phase transition}

\ShortTitle{Instantons survive ...}

\author{Falk Bruckmann\\

        Institute for Theoretical Physics, University of Regensburg, 
        D-93040 Regensburg, Germany
}

\abstract{Instantons at finite temperature consist of dyons, 
with their masses depending on the asymptotic Polyakov loop.
It is shown that the suppression of heavy dyons can explain qualitatively the behavior of 
the periodic and antiperiodic chiral condensate
and the topological susceptibility 
above the phase transition. 
}

\FullConference{8th Conference Quark Confinement and the Hadron Spectrum \\
                 September 1-6 2008\\
                 Mainz, Germany}

\begin{document}

At the Yang-Mills (here SU(2)) phase transition
the Polyakov loop starts to deviate from being traceless
and the chiral condensate drops to zero.
In addition, the chiral condensate with periodic boundary conditions survives,
the spatial string tension even grows
and the topological susceptibility decreases. 
The latter seems to indicate that topological excitations
are suppressed in the deconfined phase.

However, a correct semiclassical model at finite temperature
has to be based on dyons, visible e.g.\ as constituents of calorons
\cite{Harrington:1978ve_and_more}.
The properties of these dyons depend on the asymptotic
Polyakov loop, called holonomy, 
$\mathcal{P}_\infty\sim e^{2\pi i\omega\sigma_3}$ (we put the temperature to 1).
In particular, the two dyons have `masses', that is fractions of the action or topological charge, 
of $2\omega$ and $2\bar{\omega}\equiv1-2\omega$
in instanton units.
Hence the dyons -- selfdual or antiselfdual -- look equal for traceless holonomy,
while their masses and profiles start to differ 
for the holonomy approaching $1_2$, i.e.\ $\omega\to 0$ 
(we specialize to positive (traced) holonomies
with the negative ones available by a center transformation).
Under the conjecture, that $\tr \mathcal{P}_\infty$ equals 
the order parameter $\langle \tr \mathcal{P}\rangle$, 
the dyons are of equal mass in the confined phase,
whereas they become light  
vs.\ heavy  
in the deconfined phase.

Here we consider the simplest dyon model, namely a dilute non-interacting\footnote{
Semiclassical quantities like the fluctuation determinant, the action and the metric
\cite{Diakonov:2007nv} will generate Coulomb-like interactions at the next level of approximation.} 
gas of selfdual and antiselfdual dyons.
The effects on observables above the phase transition 
-- at least the correct tendency -- is expected to be driven by
the suppression of the heavy dyons through changing $\omega$
(a neutrality condition like in \cite{Diakonov:2007nv} is not imposed and the rest of the parameters is fixed, 
as has also been done (partly) in a semianalytic study of calorons \cite{Gerhold:2006sk}). 
For very small $\omega$ the applicability of the model is limited because the light dyons profile spreads 
and might contradict diluteness.

Firstly, according to index theorems 
and explicit calculations
at finite temperature
\cite{Callias:1977kg_and_more},
the light dyons carry a periodic zero mode each, 
while the heavy dyons carry the 
(physical) antiperiodic zero modes.
In a gas of selfdual and antiselfdual dyons the individual zero modes 
become near zero modes and eventually build up the chiral condensates.
Hence, below the phase transition the periodic and antiperiodic
chiral condensate are equal, since their carriers occur equally frequent. 
Above the phase transition, however, the physical antiperiodic
condensate should be suppressed, while the periodic one should persist 
and may even grow. 
This has indeed been observed numerically \cite{gattringer:02b_and_more} 
and is necessary in the context of dressed Polyakov loops and center symmetry \cite{bilgici:08}.

In a similar way, the surviving light dyons shall generate a larger spatial string tension, 
but a smaller topological susceptibility 
as the topological fluctuations now come in smaller units.

For a quantitative analysis one needs to complement the weight factors 
$\zeta_{l}=
\exp(-\frac{8\pi^2}{g^2}2\omega),\,
\zeta_{h}=
\exp(-\frac{8\pi^2}{g^2}2\bar{\omega})$  
by the lowest order of the metric determinant, 
which gives $d_l=8\pi\omega$ resp.\ $d_h=8\pi\bar{\omega}$ per dyon
(see \cite{Diakonov:2007nv} and references therein),
and of the fluctuation determinant, 
which yields the one-loop effective potential $P(\omega)=4\pi^2(2\omega)^2(2\bar{\omega})^2/3$,
an overall volume factor known to favor trivial holonomy \cite{Weiss:1980rj_and_more}.

Expectation values of observables in the model are computed by integrations over the dyons collective coordinates: 
three-dimensional locations and phases. 
We will, however, consider observables $O$ that only depend on the number of dyons  (and $\omega$) 
and hence the integrals give volumes,
\begin{eqnarray}
\langle O \rangle  = \frac{1}{Z}\,\exp[-V P(\omega)] 
\sum_{N_l^{\pm},\,N_h^{\pm}}
\frac{1}{N_l^+!N_l^-!N_h^+!N_h^-!}
(V\zeta_l d_l)^{N_l^++N_l^-}(V\zeta_h d_h)^{N_h^++N_h^-}\,
 O(N_l^\pm,N_h^\pm;\omega)\,,
\end{eqnarray}\vskip-0.2cm
\noindent where $N_{l,h}^\pm$ is the number of light/heavy dyons of positive/negative topological charge.
 
For the condensates we simply measure the total number of dyons 
supporting zero modes with the given boundary condition, $O_{per,aper}=N^+_{l,h}+N^-_{l,h}$.
For the topological susceptibility, the topological units are squared 
and multiplied with the number of dyons, $O_{Q^2}=(N^+_l+N^-_l)(2\omega)^2+(N^+_h+N^-_h)(2\bar{\omega})^2$.
A straightforward computation yields the following densities:
\begin{eqnarray}
 \langle O_{per} \rangle / V = 2 \zeta_l d_l\,,\quad
 \langle O_{aper} \rangle / V = 2 \zeta_h d_h\,,\quad 
 \langle O_{Q^2} \rangle / V = (2\omega)^2 \langle O_{per} \rangle / V + (2\bar{\omega})^2\langle O_{aper} \rangle / V\,.
\end{eqnarray}

In Fig.\ 1 we plot these quantities as a function of $\omega$ for different values of the coupling constant $g$.
The physical behavior, $\langle O_{aper}\rangle$ and $\langle O_{Q^2} \rangle$ suppressed 
and $\langle O_{per} \rangle$ enhanced away from the confined phase holonomy $\omega=1/4$,
is indeed found for a range of intermediate couplings $g\simeq 4$. 

\begin{figure}[!h]
 \includegraphics[width=0.32\linewidth]{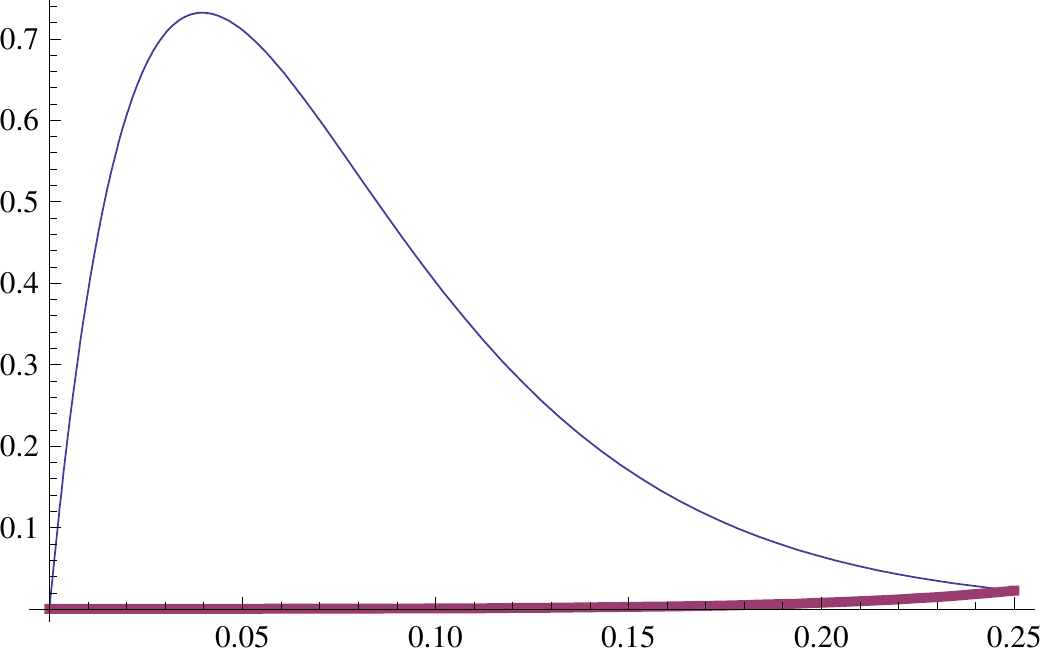}\hfill
 \includegraphics[width=0.32\linewidth]{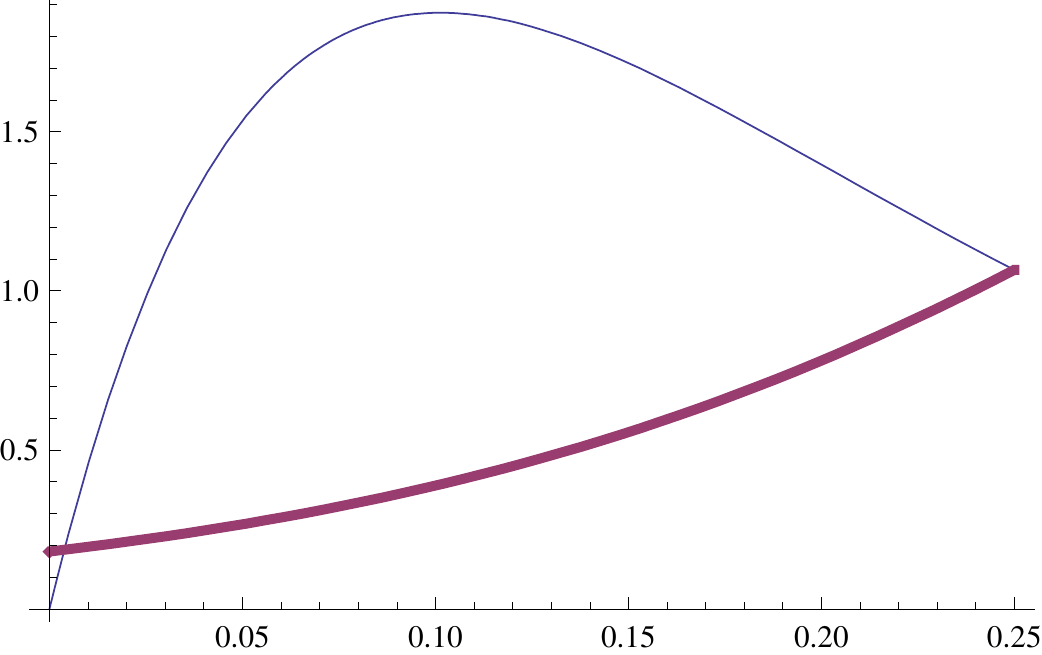}\hfill
 \includegraphics[width=0.32\linewidth]{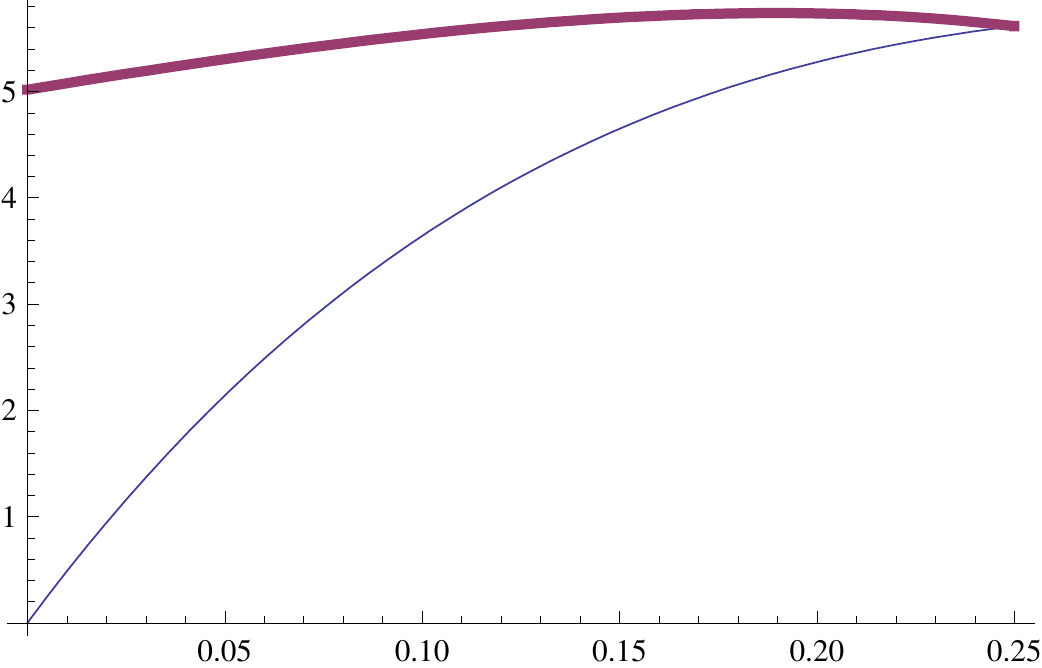}\\
 \includegraphics[width=0.32\linewidth]{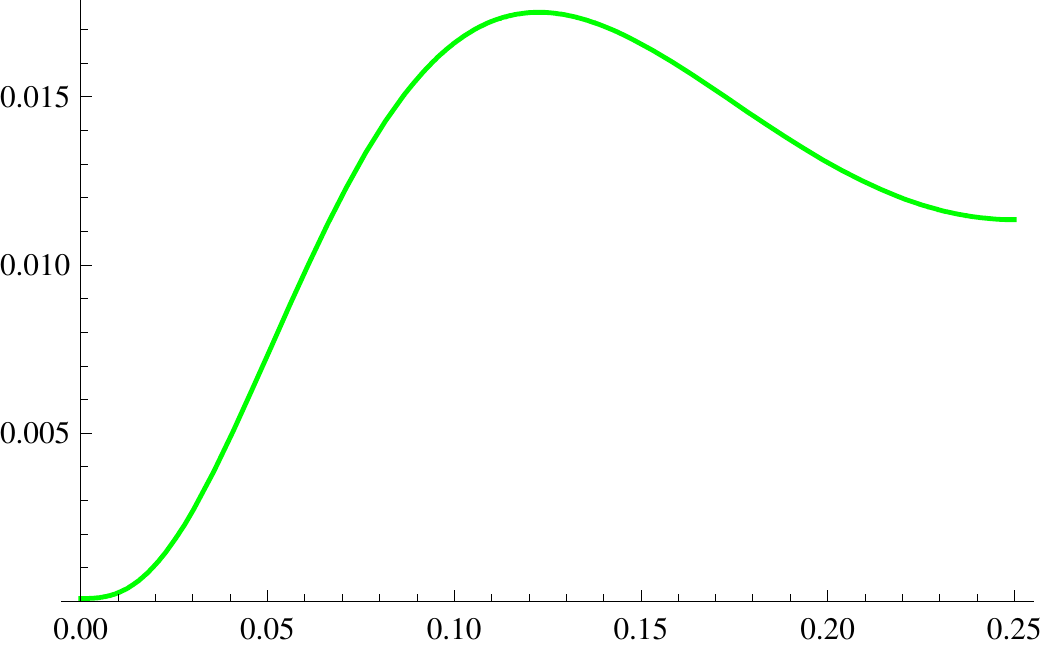}\hfill
 \includegraphics[width=0.32\linewidth]{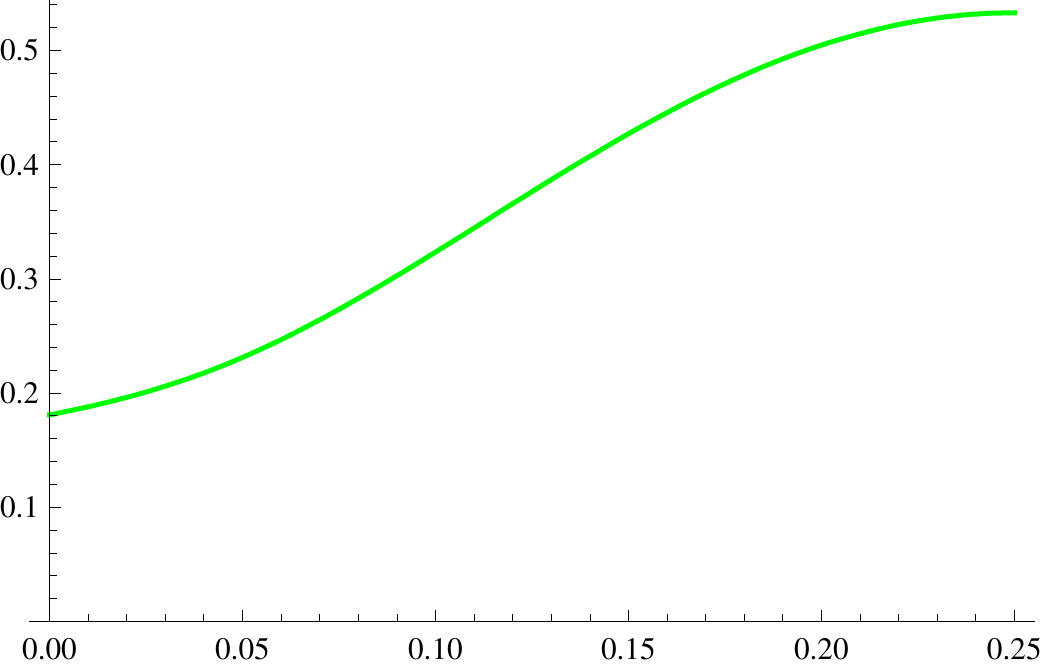}\hfill
 \includegraphics[width=0.32\linewidth]{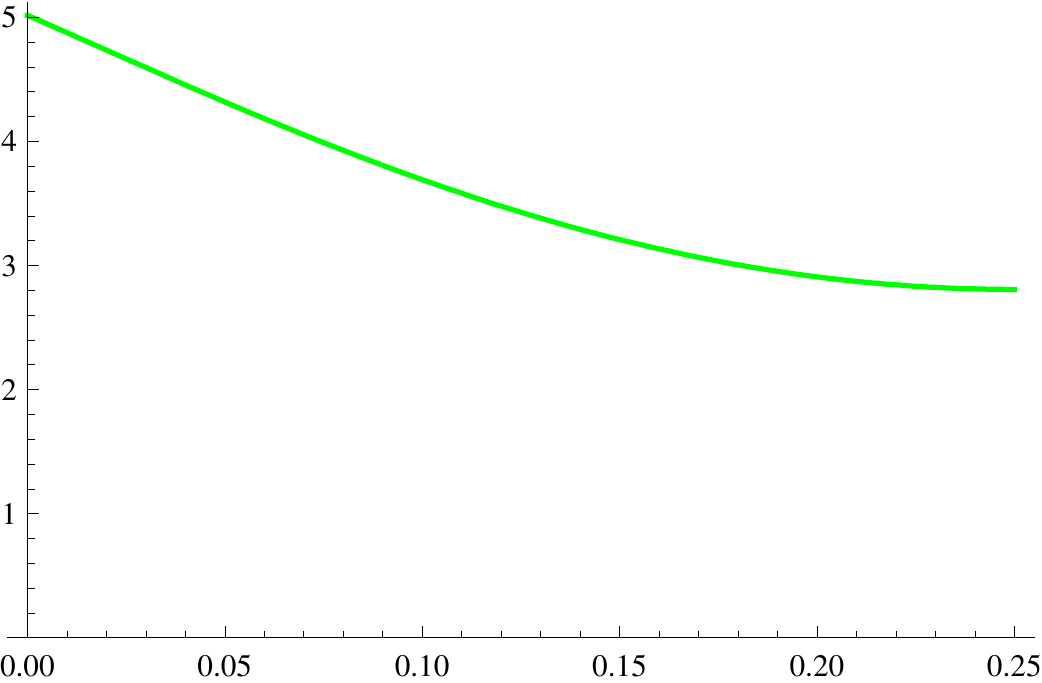}
 \caption{Expectation value of the number of periodic and antiperiodic zero modes 
(top, the latter bold) and of the topological susceptibility (bottom) as a function of $\omega$ 
for couplings $g=2.5,4,7$ (left to right).}
\end{figure}

Concerning spatial Wilson loops, it is known that individual dyons induce superconfinement
and so will our simple model
(the corresponding `string tension' may change with $\omega$ though).
For this and more precise calculations in general, a fully interacting dyon model must be investigated.\\

\noindent{\textbf{Acknowledgements}\\

\noindent The author thanks Simon Dinter, Ernst-Michael Ilgenfritz, Boris Martemyanov, Michael M\"uller-Preussker and Marc Wagner for discussions and acknowledges support by DFG (BR 2872/4-1).


\begin{thebibliography}{99}

\bibitem{Harrington:1978ve_and_more}
  B.~J.~Harrington and H.~K.~Shepard,
  Phys.\ Rev.\  D {\bf 17} (1978) 2122;
T.~C.~Kraan and P.~van Baal,
  Nucl.\ Phys.\  B {\bf 533} (1998) 627;
K.~M.~Lee and C.~h.~Lu,
  Phys.\ Rev.\  D {\bf 58} (1998) 025011;

\bibitem{Diakonov:2007nv}
  D.~Diakonov and V.~Petrov,
  Phys.\ Rev.\  D {\bf 76} (2007) 056001.

\bibitem{Gerhold:2006sk}
  P.~Gerhold, E.~M.~Ilgenfritz and M.~M\"uller-Preussker,
  Nucl.\ Phys.\  B {\bf 760} (2007) 1.

\bibitem{Callias:1977kg_and_more}
  C.~Callias,
  Commun.\ Math.\ Phys.\  {\bf 62} (1978) 213;
T.~Nye, M.~Singer, 
  J.\ Funct.\ Anal.\ {\bf 177} (2000) 203;
M.~Garcia Perez, A.~Gonzalez-Arroyo, C.~Pena and P.~van Baal,
  Phys.\ Rev.\  D {\bf 60} (1999) 031901;
E.~M.~Ilgenfritz, B.~V.~Martemyanov, M.~M\"uller-Preussker, S.~Shcheredin and A.~I.~Veselov,
  Phys.\ Rev.\  D {\bf 66} (2002) 074503;
F.~Bruckmann, D.~Nogradi and P.~van Baal,
  Nucl.\ Phys.\  B {\bf 666} (2003) 197.

\bibitem{gattringer:02b_and_more}
C.~Gattringer and S.~Schaefer,  Nucl.\ Phys.\ {\bf B654} (2003) 30;
V.~G.~Bornyakov, E.~V.~Luschevskaya, S.~M.~Morozov, M.~I.~Polikarpov, E.~M.~Ilgenfritz and M.~M\"uller-Preussker,  arxiv:
 0807.1980 [hep-lat];
  V.~G.~Bornyakov, E.~M.~Ilgenfritz, B.~V.~Martemyanov and M.~M\"uller-Preussker,
  arxiv: 0809.2142 [hep-lat];
T.~G. Kovacs, PoS {\bf LAT2008} (2008) 198;
M.~A. Stephanov,   Phys.\ Lett.\ {\bf B375}
  (1996) 249. 

\bibitem{bilgici:08}
E.~Bilgici, F.~Bruckmann, C.~Gattringer and C.~Hagen,  {\em Phys. Rev.} {\bf D77} (2008)
  094007. 

\bibitem{Weiss:1980rj_and_more}
  N.~Weiss,
  Phys.\ Rev.\  D {\bf 24} (1981) 475;
K.~Zarembo,
  Nucl.\ Phys.\  B {\bf 463} (1996) 73;
D.~Diakonov, N.~Gromov, V.~Petrov and S.~Slizovskiy,
  Phys.\ Rev.\  D {\bf 70} (2004) 036003.


\end{thebibliography}
\end{document}